\documentclass{article}
\usepackage{amsmath,graphicx,amsthm,amssymb}
\usepackage{authblk}
\usepackage[utf8]{inputenc}
\usepackage{caption}
\usepackage{todonotes}
\usepackage[a4paper, total={6.5in, 10in}]{geometry}
\usepackage{lineno}
\usepackage{setspace}
\usepackage{verbatim}
\usepackage[numbers,sort&compress]{natbib}
\usepackage{multirow}
\usepackage{array}
\usepackage{url}
\usepackage{changepage}
\usepackage{booktabs}
\usepackage{hyperref}

\title{\textbf{Coupled power generators require stability buffers in addition to inertia}}

\author[a,b,$\dagger$]{Gurupraanesh Raman}
\author[a,b,$\dagger$]{Gururaghav Raman}
\author[a,*]{Jimmy Chih-Hsien Peng}
\affil[a]{Department of Electrical and Computer Engineering, National University of Singapore, Singapore 117581}
\affil[b]{Singapore-ETH Centre, 1 CREATE Way, Singapore 138602}
\affil[*]{\footnotesize Corresponding author. E-mail:\ jpeng@nus.edu.sg}
\affil[$\dagger$]{\footnotesize Equally contributing authors}

\date{}

\begin{document}
\maketitle

\section*{}
\begin{adjustwidth}{1.6cm}{1.6cm}
\hspace*{5.8cm}\fontsize{12}{12}\selectfont{\textbf{Abstract}}\vspace*{0.4cm} 

Increasing the inertia is widely considered to be the solution to resolving unstable interactions between coupled oscillators. In power grids, Virtual Synchronous Generators (VSGs) are proposed to compensate the reducing inertia as rotating synchronous generators are being phased out. Yet, modeling how VSGs and rotating generators simultaneously contribute energy and inertia, we surprisingly find that instabilities of a small-signal nature could arise despite fairly high system inertia. Importantly, we show there exist both an optimal and a maximum number of such VSGs that can be safely supported, a previously unknown result directly useful for power utilities in long-term planning and prosumer contracting. Meanwhile, to resolve instabilities in the short term, we argue that the new market should include another commodity that we call stability storage, whereby---analogous to energy storage buffering energy imbalances---VSGs act as decentralized stability buffers. While demonstrating the effectiveness of this concept for a wide range of energy futures, we provide policymakers and utilities with a roadmap towards achieving a 100\% renewable grid.
\end{adjustwidth}

\section*{Significance Statement}
As renewables increasingly replace rotating synchronous generators in the power grid, its inertia reduces, jeopardizing the self-synchronizing property of the generators. Utilities are therefore exploring the provision of inertia from the renewables themselves, designating a few as virtual synchronous generators. Modeling such a future, we surprisingly find an upper limit on the number of generators that can participate in providing inertia. Rather, stability buffers are necessary as more renewables are added, or owing to short-term network changes---similar to how energy storage buffers generation-demand imbalances. Our results have direct policy implications on long-term planning and market design for future sustainable grids, and more broadly, add to the literature on the stability of coupled dynamic systems.

\medskip

\section*{Introduction}
Power grids around the world are transitioning from conventional fossil-fuel-based generation to renewables such as solar photovoltaic and wind generation in a bid to combat climate change.
This has ushered in an era where the dominant share of electricity generation may no longer come from large rotating synchronous machines, but rather, smaller power electronic converters---called inverters---that transfer DC power from the renewables into the AC power grid.
A key consequence of replacing synchronous generation by inverters is the reduction in the grid inertia, which has heretofore existed in the form of the inherent kinetic energy in the spinning rotors~\cite{Yang2018burden, NREL_inertia}.
When inertia becomes insufficient, small power disturbances in the power grid result in larger frequency deviations, which when exceeding allowable limits, can trip generators and cause blackouts~\cite{Uriarte2015microgrid}.
Cognizant of this, many grid operators around the world (e.g., Hawaii~\cite{GFMroadmap}, Puerto Rico~\cite{GFMroadmap}, United Kingdom~\cite{VSM_NationalGrid}, Europe~\cite{GFMentsoe}) find that inverters need to start contributing inertia in the near future, considering different energy futures and approaches to decarbonization within and across nations.
Such inverters would be different from the overwhelming majority of inverters in the present that simply inject renewable power to support the grid (i.e., operate in the \textit{grid-following mode}): they would actively control the voltage and frequency of the grid (i.e., operate in the \textit{grid-forming mode}).
Effectively, they would operate as Virtual Synchronous Generators (VSGs) to emulate the inertial dynamics of their rotating counterparts, and are planned to be incentivized for this service through a new stability market (e.g., see ref.~\cite{Stability_market_NationalGrid} for the UK).

As yet, VSGs have not been implemented in bulk power grids at any scale, and there are no operational standards as to how systems with both rotating and virtual synchronous generators should be operated, especially as their relative shares change in the future.
Motivated by this lack of industry experience with VSGs and given their imminent introduction in grids around the world, we ask the question: is maintaining adequate inertia alone sufficient for maintaining the stability of the grid, for the entire range of expected energy futures?
In studying the stability implications of an inertia market, we model how energy will be shared amongst the synchronous generators and the VSGs.
The manner in which they do this and respond simultaneously in real time governs the system's \textit{small-signal stability}, a critical measure of how quickly incremental power disturbances are damped out.
The small-signal stability is influenced strongly by the control parameters of the generators (rotating or virtual), as well as the inertia.
Nevertheless, there are no comprehensive studies in the literature focusing on the nexus between inertia, energy sharing amongst the various generators, and small-signal stability.
Such a study is in fact essential as VSGs are introduced in the grid due to two factors.
First, VSGs are operated by prosumers---entities who consume and produce electricity at the same time---who may flexibly connect, disconnect, or operate their inverters to satisfy their own financial ends in a deregulated environment rather than being concerned with global stability, which they have no awareness of.
Second, the inverters are located at the grid edge rather than in a few centralized locations as the synchronous generators were, rendering real-time control unscalable~\cite{Zhang2016interactive}.

With this in mind, we take the power grid of the Greater London metropolis, and analyze its small-signal stability for various energy futures forecast by the grid operator, National Grid.
Surprisingly, we demonstrate that the grid can become unstable even at high inertia levels and low shares of inverter generation when the VSGs are not properly sized.
We further show that for a given inertia level, there exists an optimal as well as a maximum number of VSGs that can be incorporated safely into the grid.
In doing so, we effectively provide a long-term roadmap for energy policymakers and utilities in planning the required inertia, and the number and capacity of the VSGs for each energy future.
While the results shown here are for the Greater London network, the trends are explained by control-theoretic concepts and are therefore applicable to other grids as well.
Next, we show that to guarantee the stability in the short term, the VSGs themselves must necessarily participate in stabilizing the grid and that external interventions are not successful.
To realize this, we view small-signal stability as a decentralized and additive resource that can be commoditized through what we call \emph{stability storage}, so that it can also be traded in the stability market.
Drawing inspiration from the well-known use of energy storage to mitigate generation-demand imbalances in a decentralized manner~\cite{Stephan2016limiting}, we demonstrate that having stability storage can effectively solve the instability problem.

Broadly, VSGs have the ability to enable grids with 100\% renewable generation, and are crucial for achieving a sustainable electric grid in nations that do not have appreciable hydro or geothermal potential.
Therefore, although our study appears to arise from a control perspective, the answers---which define the operational philosophy of future power grids worldwide---govern the smooth transition towards sustainability.

\section*{Related works}
AC power grids with synchronized generators are complex systems exhibiting a variety of instability phenomena that are of interest to both power engineers~\cite{Gao2017research, aderibole2017domain, Nikolakakos2014enhancement, Uriarte2015microgrid, pagnier2019inertia} and network scientists~\cite{menck2014dead, simpson2016voltage, motter2013spontaneous, tyloo2019key, molnar2021asymmetry, hellmann2020network, dorfler2013synchronization}.
Specific to the oscillatory dynamics of generators presently under consideration, our work differs from previous studies in two substantive ways.
First, previous studies~\cite{motter2013spontaneous, molnar2021asymmetry, aderibole2017domain, Nikolakakos2014enhancement, Uriarte2015microgrid} have examined the impact of generator controller parameters and network interconnections on the ability of generators to self-synchronize, in other words, to preserve small-signal stability in the grid.
In particular, the positive impact of higher inertia on the small-signal stability has been reported in ref.~\cite{pagnier2019inertia, motter2013spontaneous,tyloo2019key}, highlighting the need for VSGs in future power grids.
Nevertheless, an unanswered question remains as to how many inverters should participate as VSGs---contributing both inertia as well as power---and what (if any) is the limit on this number. 
Considerations of inertia and power sharing, when combined, present a challenge as there is no degree of freedom in the primary controller parameters whereby they can be tuned to improve stability; see Methods.
Second, common to the prior studies is that they have treated small-signal stability as a global measure, and not as a distributed commodity that can be provided by multiple grid-edge devices in an additive manner. 
As such, small-signal stability has never been commoditized for provision through a market, let alone in the context of VSGs.
Our paper demonstrates how this can be done using virtual impedance control, a well known stabilizing strategy~\cite{he2011analysis, wang2014virtual, guan2015new, huang2019plug}.

There is another body of literature focused on developing control solutions for grid-forming inverters that result in grids that are theoretically stable for any configuration, e.g., by using lead-lag compensators~\cite{Raman2020mitigating, Dheer2019improvement}, measurement frame rotation~\cite{de2007voltage}, etc.
However, such stability guarantees do not hold when synchronous generators and inverters are both present, or when the latter have heterogeneous control strategies~\cite{raman2020optimal}.
Given that in practice, inverters and synchronous generators are likely to co-exist, and that inverters are likely to come from different vendors and have different operational constraints specified by their individual prosumer operators~\cite{Parag2016electricity}, these control solutions may not sufficiently ensure the robustness of the future grid.
Centralized coordination, however, is not a scalable alternative given the number and dispersed nature of the generators~\cite{babazadeh2018robust}; an alternate decentralized market-based approach is proposed here.
We mention that while our focus is only on grid-forming inverters, some energy markets presently do allow grid-following inverters to provide inertia (called fast frequency response). 
However, we do not consider these in our analyses due to three reasons: (i)~their response time is too slow for this service alone to be adequate without grid-forming inverters~\cite{Alkez2020critical,NREL_inertia}; (ii)~they may suffer from an energy-recovery phase that lowers the frequency again~\cite{Alkez2020critical}; and (iii)~such grid-following inverters do not affect the small-signal stability of the grid~\cite{Bottrell2013dynamic}.

\begin{figure}[tp]
	\centering
	\includegraphics[width=\columnwidth]{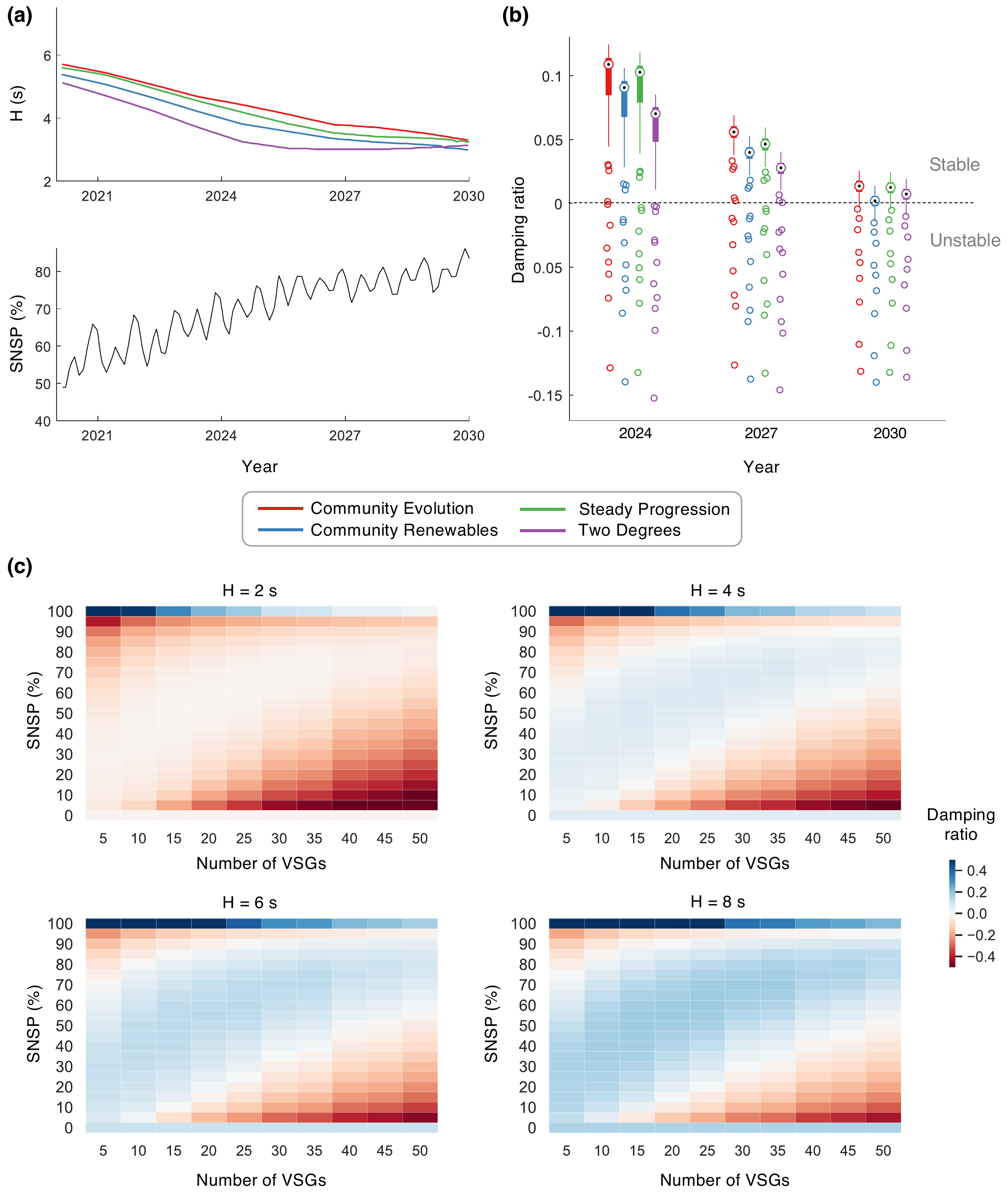}
	\caption{\textbf{Small-signal instability in the Greater London power grid for various energy futures considering virtual synchronous generators.} 
	\textbf{(a)}~Projections of the mean system inertia constant, $H$, for four future energy scenarios and the system non-synchronous penetration (SNSP) (Source: National Grid~\cite{Inertia_futures,Gomersall_NationalGrid}).
	\textbf{(b)}~Distributions of the damping ratio for the projected grid in 2024, 2027, and 2030. We assume 30 VSGs provide energy and inertia in addition to the currently-existing rotating synchronous generators. Results are shown for 100 simulations where the locations of the inverters were chosen randomly. The central mark, bottom, and top edges of each box plot represent the median, 25th, and 75th percentiles, respectively.
	\textbf{(c)}~Impact on the grid stability as the number of VSGs and SNSP change. Each cell in the heat maps is the average damping ratio over 100 simulations where the locations of the inverters were chosen randomly.}
	\label{fig:Heatmap_variation_fossil_fuel_dependence}
\end{figure}

\section*{Results}

\subsection*{Inertia -- small-signal stability nexus}
To evaluate the stability of the power grid when VSGs provide power and inertia in tandem with rotating generators, we use the distribution grid of Greater London as a case study. 
We consider four future energy scenarios developed by the system operator, National Grid, to represent the range of credible futures for the UK energy system to achieve full decarbonization by 2025~\cite{FES2019_nationalgrid}.
These scenarios include projections~\cite{Inertia_futures} of the system inertia, which are shown in Fig.~\ref{fig:Heatmap_variation_fossil_fuel_dependence}(a).
Additionally, we obtain projections for the average System Non-Synchronous Penetration (SNSP)~\cite{Gomersall_NationalGrid}, which is the overall share of the power supplied by inverter-based generators.
Finally, we derive the grid topology using publicly available data, simulate the system for different inertia and SNSP projections, and examine the small-signal stability at each step; see Methods.
The latter is quantified by the damping ratio of the least stable non-zero eigenmode, a fraction that varies between $[-1,1]$---the grid is more stable if it is positive and closer to 1, and more unstable if it is negative and closer to -1.
In practice, unstable grids experience generator and/or line trips from exceeding the voltage and/or frequency limits, resulting in blackouts.

The results of our analysis are summarized in Fig.~\ref{fig:Heatmap_variation_fossil_fuel_dependence}(b) for three specific instances in the future.
Here, we randomly vary the inverters' locations in the grid for each projected inertia and SNSP value, and depict the distribution of the resultant damping ratios in the box plot.
Our results show that instabilities may arise when VSGs are introduced regardless of the future energy scenario---with the damping ratio worsening with reducing system inertia---underscoring the need to consider small-signal stability while designing pathways for the sustainability transition.
We further study how the number of VSGs in the grid---a parameter whose value is uncertain in the future and could vary in real-time---impacts its stability, and whether increasing the overall system inertia can guarantee stability.
Referring to Fig.~\ref{fig:Heatmap_variation_fossil_fuel_dependence}(c), two key observations can be gleaned from our simulations.
First, the stability of the grid improves when the overall inertia is increased. 
Surprisingly however, we find that even with a fairly high inertia constant, say $H=8s$, the grid proves to be unstable for certain values of SNSP and number of VSGs, indicating that simply satisfying a minimum inertia criterion would be insufficient to guarantee the stability of the future power grid. 
Second, there exists a symmetry in each of the heat maps depicted in Fig.~\ref{fig:Heatmap_variation_fossil_fuel_dependence}(c)---nearly along the anti-diagonal---resulting in two distinct regions of low stability.
On closer examination, we find that instabilities worsen when there is a larger disparity between the capacities and therefore the droop gains of the different generators in the grid, be it virtual or rotating (see Methods).
In the top-left of the heat maps, the latter contribute a very small fraction of the total generation capacity, and consequently, have high droop gains. 
In contrast, in the bottom-right corner, VSGs supply the smaller fraction, and have high droop gains. 
These high-droop-sources exhibit more aggressive control actions, and consequently, unstable behavior.
Further, in the bottom-right, the stability worsens as the number of VSGs increases. 
The addition of more inverters for the same SNSP value not only increases their individual droop gains, but also reduces the effective electrical distance between any pair of inverters; this increases the feedback gain for the inverters' controllers, further reducing the stability.
In the first and last rows of the heat maps, the higher-droop-sources are entirely eliminated, resulting in higher stability and the discontinuity seen in the damping trends.
Commenting on the above observations, the symmetry along the anti-diagonal is reflective of the similarity between VSGs and rotating synchronous generators. 
Given the fact that the former were developed precisely to mimic the latter, there is no qualitative distinction between the two as far as the grid stability is concerned.
Overall, our results show that for every inertia and SNSP value, there exists an upper limit on the number of VSGs that can stably coexist in the system---this limit increases with the inertia.
For example, assuming an inertia constant of 4~s and SNSP=50\%, we find that the Greater London grid can support, at a maximum, 35 VSGs, and for 6~s, 40 VSGs.
When the number of VSGs is further increased, the bottom instability region takes over the entire SNSP range at some point; this limit also increases with inertia.
The above observations have implications for selecting the optimal number of grid-forming inverters for varying power systems and energy futures.

\begin{figure}[ht]
	\centering
	\includegraphics[width=0.9\columnwidth]{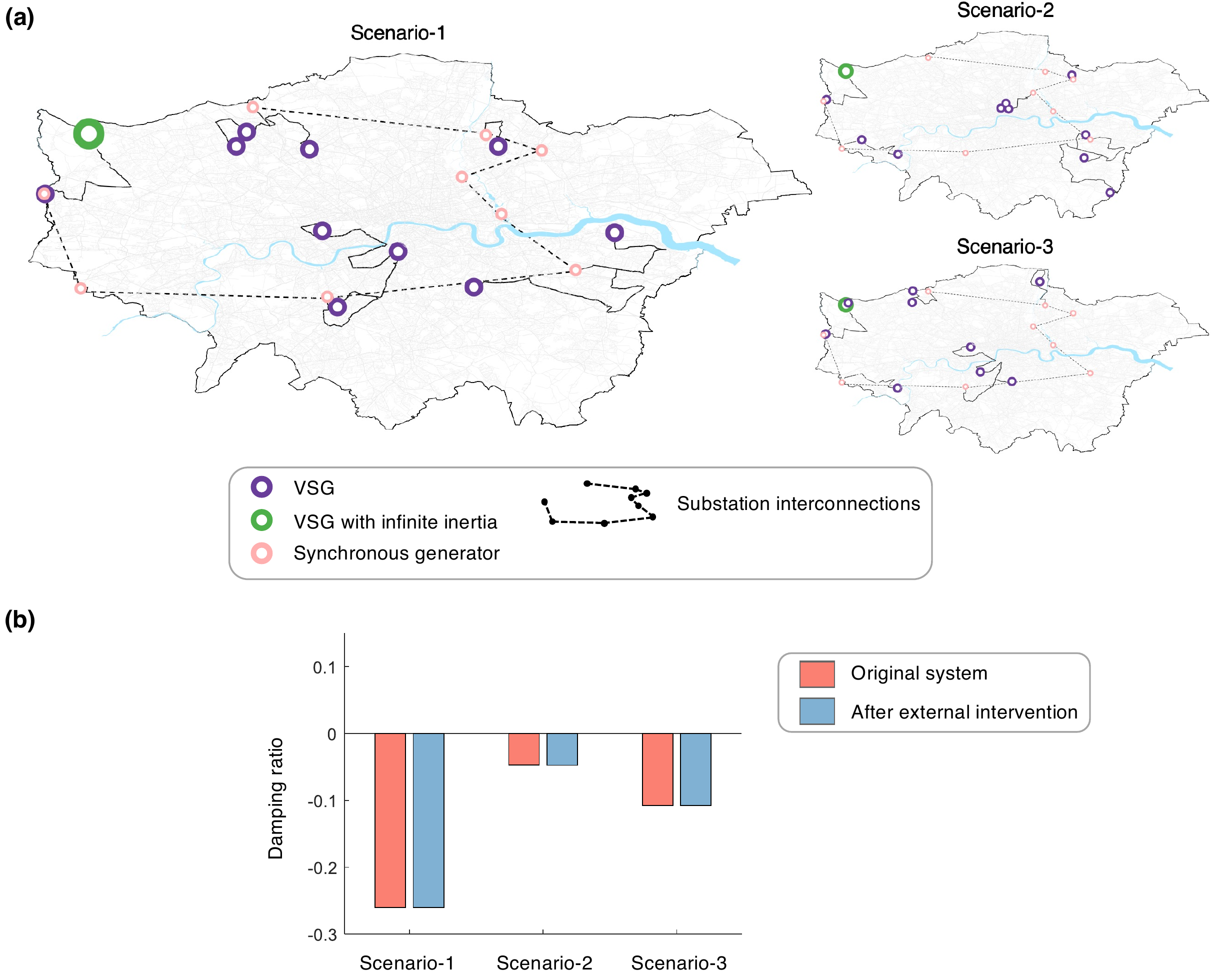}
	\caption{\textbf{Demonstrating how external interventions may not guarantee stability of the future grid.} 
	\textbf{(a)}~The figure presents three scenarios where 10 VSGs share the load with the existing rotating synchronous generators. In each scenario, a new VSG, marked in green, is added to provide infinite inertia in an exaggerated attempt to stabilize the grid.
	\textbf{(b)}~Damping ratios of the scenarios in (a) before and after the external intervention, showing no improvement in stability.}
	\label{fig:stable_unstable_scenarios}
\end{figure}

\subsection*{External interventions for stability improvement}
Presently, some power grids utilize external stabilizing grid-following converters to damp out oscillations, e.g., using wind farms~\cite{Wang2015control}.
Here, we apply the same principle for the VSG grid under study, and examine whether an additional VSG can be added to the grid to stabilize it.
To this end, we consider three randomly chosen unstable scenarios for the Greater London grid---see Fig.~\ref{fig:stable_unstable_scenarios}(a)---where in each, there are 10 VSGs in addition to the currently existing synchronous generators.
The respective damping ratios are shown in Fig.~\ref{fig:stable_unstable_scenarios}(b).
Intuitively, a newly-added inverter can provide the highest possible stability if it behaves as an \textit{infinite bus} with zero droop coefficients, contributing infinite inertia~\cite{Mohamed2008adaptive}. (This would also be mathematically identical to adding a high-capacity rotating synchronous generator instead.)
On performing the stability analysis with the additional VSG, we find that the damping ratio for each scenario either remains the same or slightly decreases.
Given that an infinite bus---arguably the most stable VSG configuration---cannot mitigate the instability, our results indicate that external interventions such as the addition of new stabilizing generators may not be beneficial to improve the small-signal stability.
Instead, we hypothesize that the instability can be successfully mitigated if the existing inverters themselves participate in the stabilization process.
Preferably, they would respond in a \emph{decentralized} manner, i.e., only use information available locally and not rely on real-time instruction from the grid operator; this would enable the control philosophy to be scalable to larger grids.

\subsection*{Stability storage analogue for energy storage}
\begin{figure}[htbp]
	\centering
	\includegraphics[width=0.93\columnwidth]{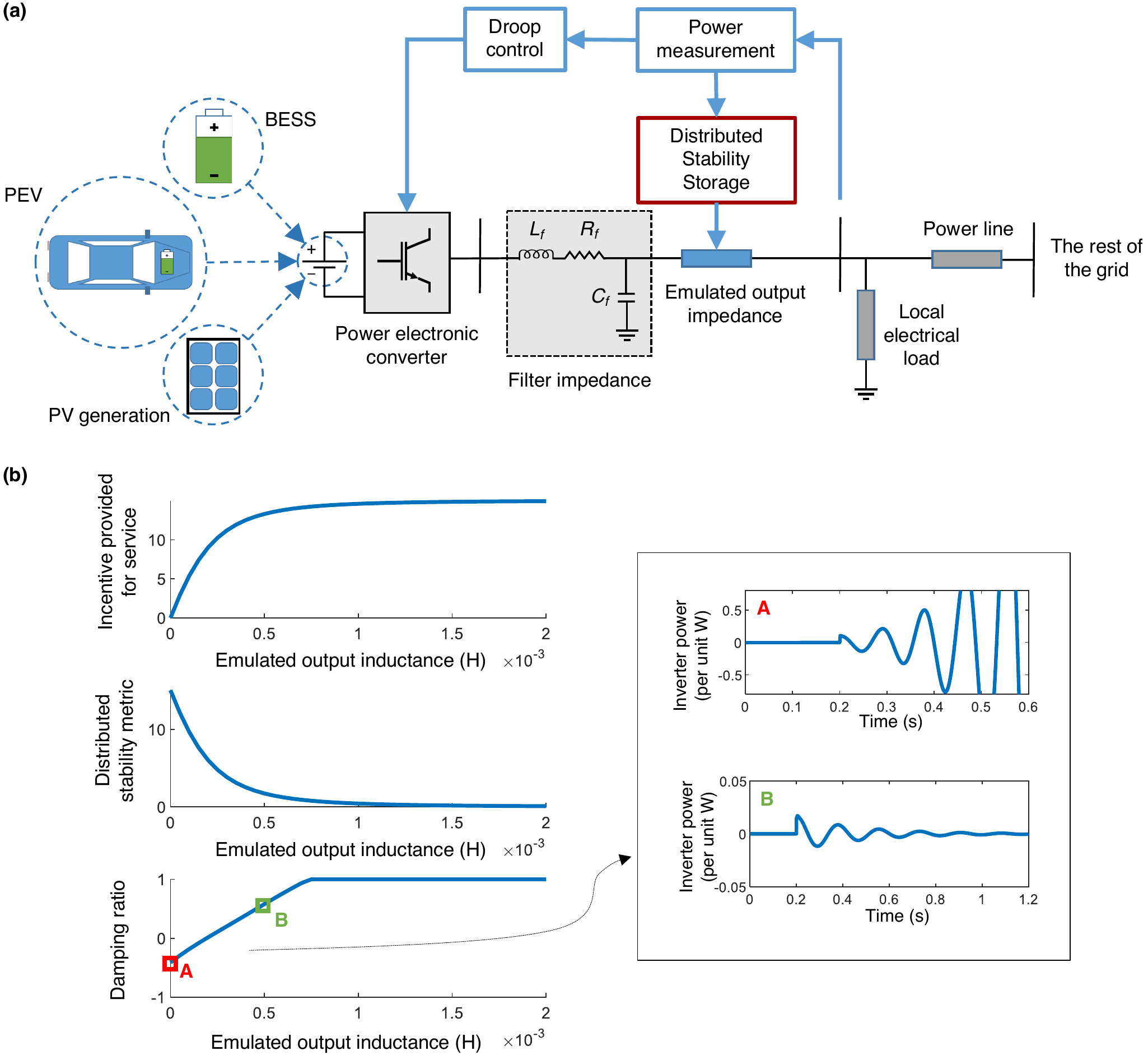}
	\caption{\textbf{How VSGs can stabilize the power grid by performing as what we call \textit{stability storage}.} 
	\textbf{(a)}~Schematic of a VSG connected to the grid. The energy source could be a battery, electric vehicle, or renewable generation. The VSG emulates a virtual impedance to contribute stability storage to the grid.
	\textbf{(b)}~Illustrating how the proposed distributed stability metric and damping ratio of the VSG vary as the emulated output inductance increases. The incentive for providing stability storage is simply the negative of the distributed stability metric, offset to be always positive. The insets on the right present time responses of the system to a small load disturbance at 0.2~s.
	[BESS: Battery Energy Storage System, PEV: Plug-in Electric Vehicle, PV: Photovoltaic] }
	\label{fig:FDC_working}
\end{figure}

We have already shown that the stability is affected by the network topology, generators' locations, sizes, and primary controller parameters (inertia constant and damping factor for rotating synchronous generators, and analogously droop gains and power filter time constant for VSGs; see Methods).
Of these, generator locations and sizes are generally fixed for a particular system, and for synchronous generators, as are the primary controller parameters (although power system stabilizers~\cite{Ghosh1984power} can be enabled if existing, to improve the damping factor). 
For VSGs, the inertia and energy sharing requirements fully determine the droop gains and power filter time constant.
Therefore, the only parameters that can be potentially changed are the network impedances that interconnect the various generators.
Intuitively, a solution to the instability problem is to increase the impedance between the inverters, as this reduces the feedback gain. 
While the physical power lines connecting the inverters cannot be changed in practice, one can however change the \emph{effective impedance} between them by suitably modifying the inverter control scheme.
More specifically, this could be achieved by the inverters emulating an additional impedance at their outputs---see Fig.~\ref{fig:FDC_working}(a)---which is popularly called virtual impedance control in the literature~\cite{he2011analysis, wang2014virtual}.

\begin{table}[htp]
\centering
\caption{Analogy between energy storage and stability storage.}
\label{tab:analogy}
\begin{tabular}{p{3cm} p{6cm} p{6cm}}
    \toprule
     & \textbf{Energy storage} & \textbf{Stability storage} \\
    \midrule
    Capability & Bridge the generation--demand gap due to intermittency of stochastic generation. & Contribute additional impedance to maintain stability amongst interacting generators.\\
    \midrule
    Rewarded quantity & Power capacity and energy contributed. & Improvement in the DSM.\\
    \midrule
    Nature of control & Decentralized, can be provided as a distributed service. & Decentralized, can be provided as a distributed service. \\
    \midrule
    Potential participants & Any prosumer with a battery or other storage connected to the grid. & Any prosumer operating a VSG.\\
\bottomrule
\end{tabular}
\end{table}

In this paper, we develop an adaptive mechanism for exploiting the virtual impedance control strategy in a decentralized fashion through the stability market where inertia is also traded.
Central to this is to view the output impedance of a VSG as \emph{stored stability} in the inverter, which may be contributed to the grid when needed.
This is directly analogous to how energy storage devices provide stored energy when it is not available from other sources, e.g., intermittent renewable generation; see Table~\ref{tab:analogy}.
Energy storage devices function by providing power, within their capacity, to the grid until the generation deficit is compensated fully.
Equivalently, a stability storage device should contribute output impedance until it compensates for the instability introduced by the addition of inverter(s) to the grid. 

The challenge remains to quantify the contribution of each stability storage device and the corresponding incentive due to it.
While an energy storage device is compensated for the amount of the instantaneous power capacity and overall energy that it provides, in the stability analogue, the contribution of each participant is interlinked and not immediately separable as the overall system stability (quantified by the damping ratio) is a global property. 
Therefore, we propose a metric, termed as \emph{distributed stability metric} (DSM) that can capture the individual contribution of each VSG to the overall stability, see Fig.~\ref{fig:FDC_working}(b) and Methods.
Note that different nodes have different interconnection impedances and VSGs with different controller parameters, and as a result, contribute varying amounts to the (in)stability.
When a new inverter is added, the DSM quantifies how much instability this addition contributes to the grid, and also how much \emph{additional} stability storage must be added to maintain stability.
The latter can be contributed by a single inverter (e.g., the newly-added inverter), and/or by other existing inverters.
The incentive offered for this grid service is directly proportional to the contributed stability, i.e., the change in the DSM before and after an inverter participates in this service. 
Referring to Fig.~\ref{fig:FDC_working}(b), by the very design of the DSM, there is an upper limit on the improvement in stability versus the emulated impedance; this is in turn reflected in the incentive as well, see Methods.
In practice, the DSM can be calculated for each inverter using only locally measurable parameters and reported to the grid operator for obtaining monetary compensation for the service, similar to energy metering.
Multiple stability storage devices can participate in a decentralized manner simultaneously; this additivity is important to giving prosumers the freedom whether to participate in this grid service or not.
Ultimately, the goal is to delegate and empower prosumers that benefit from the providing energy and inertia to the grid to also be responsible for its small-signal stability.

\begin{figure}[tp]
	\centering
	\includegraphics[width=\columnwidth]{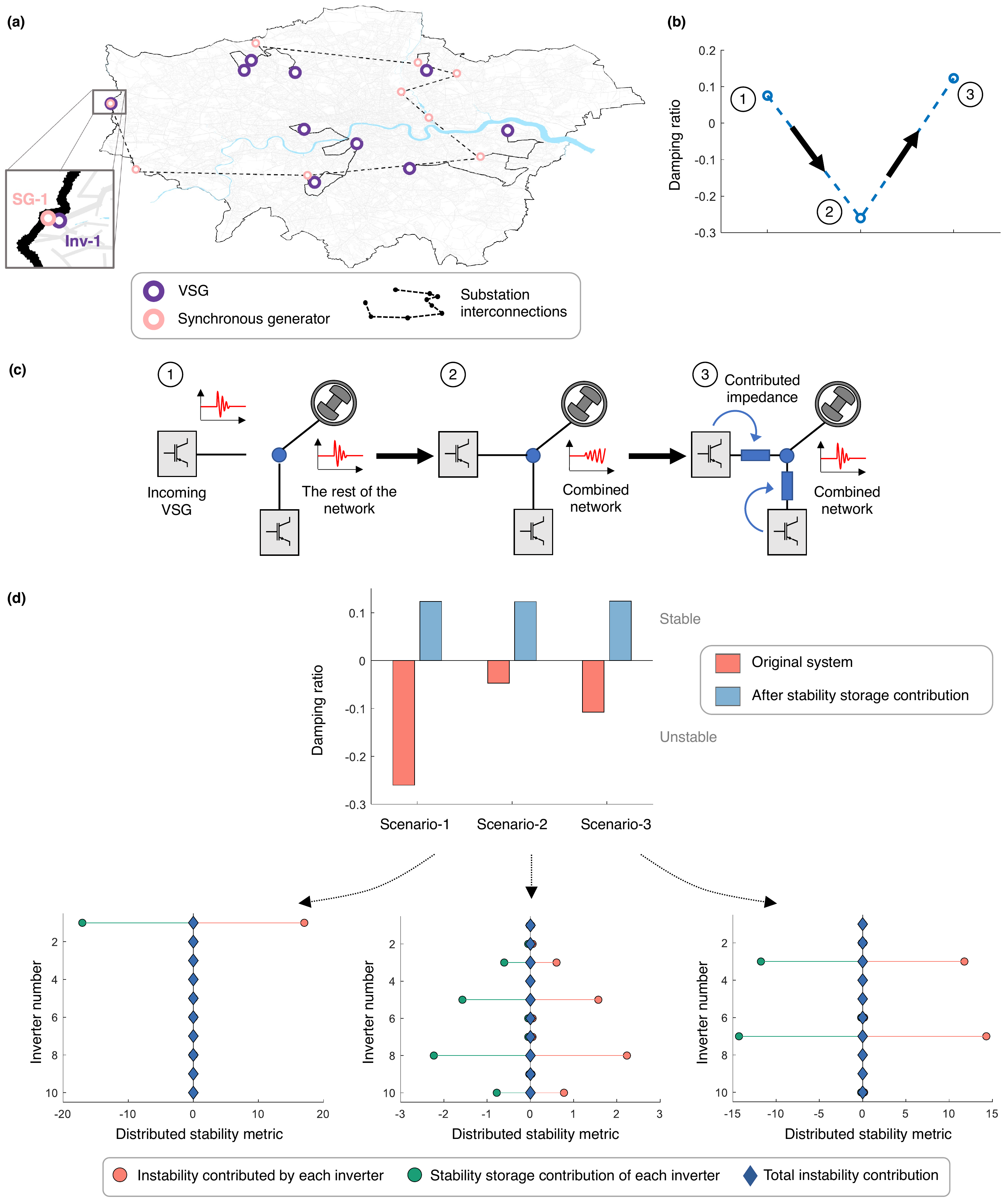}
	\caption{\textbf{Using VSGs as stability stores to expand inverter-based generation in the power grid.}
	\textbf{(a)}~Location of the rotating and virtual synchronous generators in Scenario-1 from Fig.~\ref{fig:stable_unstable_scenarios}, showing the addition of a new VSG, named Inv-1.
	\textbf{(b)}~Damping ratio in (a) before and after the addition of Inv-1. Also shown is the damping ratio after all VSGs contribute stability storage.
	\textbf{(c)}~Illustrating how a new VSG can be added to the grid while preserving stability. The following sequence is depicted: an incoming VSG is to be added to a stable grid; the combined network becomes unstable as a result; and the combined network regains its stability after the VSGs participate as stability stores.
	\textbf{(d)}~Damping ratio for the three unstable scenarios in Fig.~\ref{fig:stable_unstable_scenarios} before and after all inverters provide stability storage. Also shown is the contribution of each inverter to the instability of the grid, and the effect of stability storage in mitigating the respective instability contributions. The incentive provided to each inverter for its service is proportional to the stability storage contributed.
	}
	\label{fig:Solving_instability_LondonGrid}
\end{figure}

\subsection*{Leveraging stability storage for stabilizing VSGs}
We now take up the three unstable scenarios from Fig.~\ref{fig:stable_unstable_scenarios}(a) and demonstrate how the stability storage functionality stabilizes them.
We begin with Scenario-1.
While the system is unstable, interestingly, we find that the instability occurs only when one particular VSG, marked as `Inv-1' in Fig.~\ref{fig:Solving_instability_LondonGrid}(a), is present in the grid; see Fig.~\ref{fig:Solving_instability_LondonGrid}(b) and (c).
This is in part due to its close proximity to a neighboring synchronous generator `SG-1'.
We now demonstrate how the proposed stability storage service enables Inv-1 to be stably added to the grid.
Referring to Fig.~\ref{fig:Solving_instability_LondonGrid}(b), the system damping falls from $0.0756$ to $-0.2602<0$ when Inv-1 is connected; this damping change can be detected locally by the inverters through power measurements. 
Subsequently, all inverters, including the newly added one, contribute additional output impedance, what we call stability storage.
The particular contributions of each inverter vary according to their location, but the goal is to bring each DSM below a desired threshold; see Methods and Fig.~\ref{fig:Solving_instability_LondonGrid}(d).
With stability storage, the damping ratio becomes $0.1232>0$, or, the grid becomes stable.
We find similar outcomes for Scenarios-2 and -3 as well.

\begin{figure}[tp]
	\centering
	\includegraphics[width=\columnwidth]{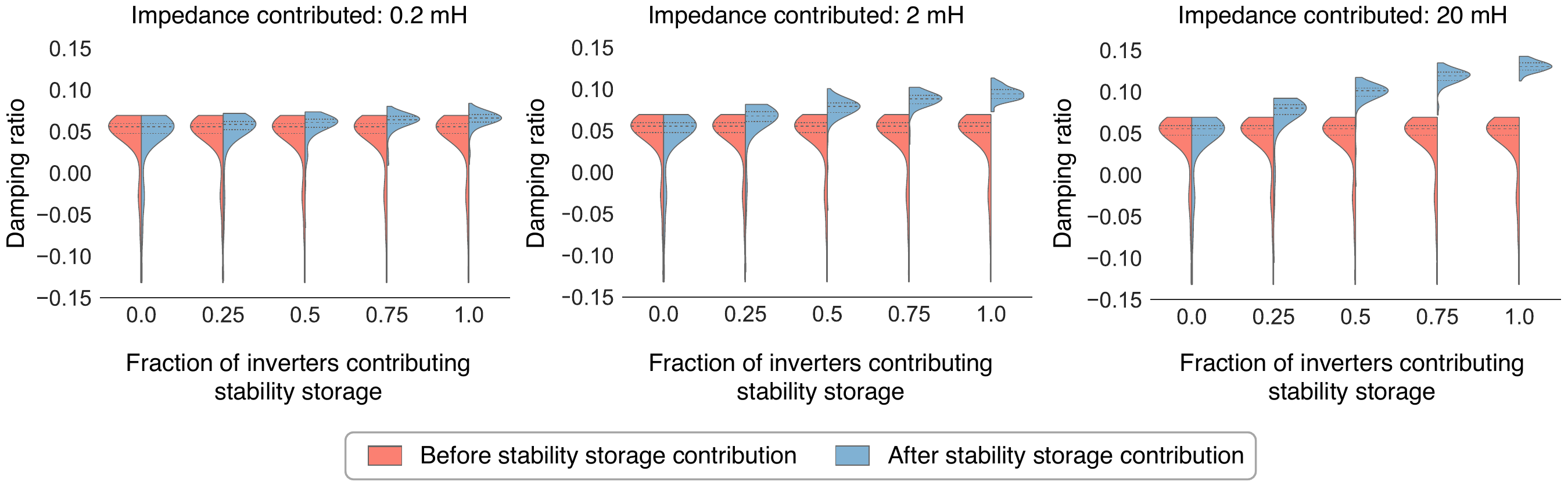}
	\caption{\textbf{Varying the VSG participation in the stability storage service.}
	Damping ratio before and after stability storage contribution with a varying fraction of inverters participating in this service. Results are presented over 100 simulations in which there are 30 VSGs randomly located across the network. Three results are shown when the impedance contribution of each inverter varies as 0.2~mH, 2~mH, and 20~mH. 
    The system inertia constant $H$ is maintained at 4~s, with SNSP set at 80\%. 
	Each pod of the violin plot depicts the quartiles of the respective distributions.}
	\label{fig:violin-plots}
\end{figure}

Next, in cognizance of the likelihood that not all prosumers/inverters would be willing to participate in the stability storage service, we vary the fraction of inverters that do and analyze the stability of the Greater London grid considering 30 VSGs.
For the purposes of this analysis, we assume that each participating VSG contributes the same amount of output impedance to the grid.
The distribution of the damping ratios obtained over 100 simulations---where the locations of the inverters change randomly---are presented in Fig.~\ref{fig:violin-plots}.
(SI Appendix, Note~1 presents the same results for different values for the number of VSGs, inertia, and SNSP.)
Here, we make two observations:
(i)~the improvement in the stability is higher when more VSGs participate as stability stores, and when the magnitude of impedance contributed by each is higher; and
(ii)~the more the penetration of stability storage, the lower the impedance needed to ensure stability across a wide range of scenarios.
These behaviors can be attributed not only to the additive property of stability storage, but also the fact that certain inverters contribute more towards instability than others (e.g., see Inv-1 in the case study from Fig.~\ref{fig:Solving_instability_LondonGrid}(a)).
Therefore, stored stability available from or near these inverters is more impactful in recovering the damping than those farther away.
As such, the more inverters that provide the stability storage service, the more the chance of an opportune inverter being available to provide much-needed output impedance. 

\begin{figure}[tp]
	\centering
	\includegraphics[width=\columnwidth]{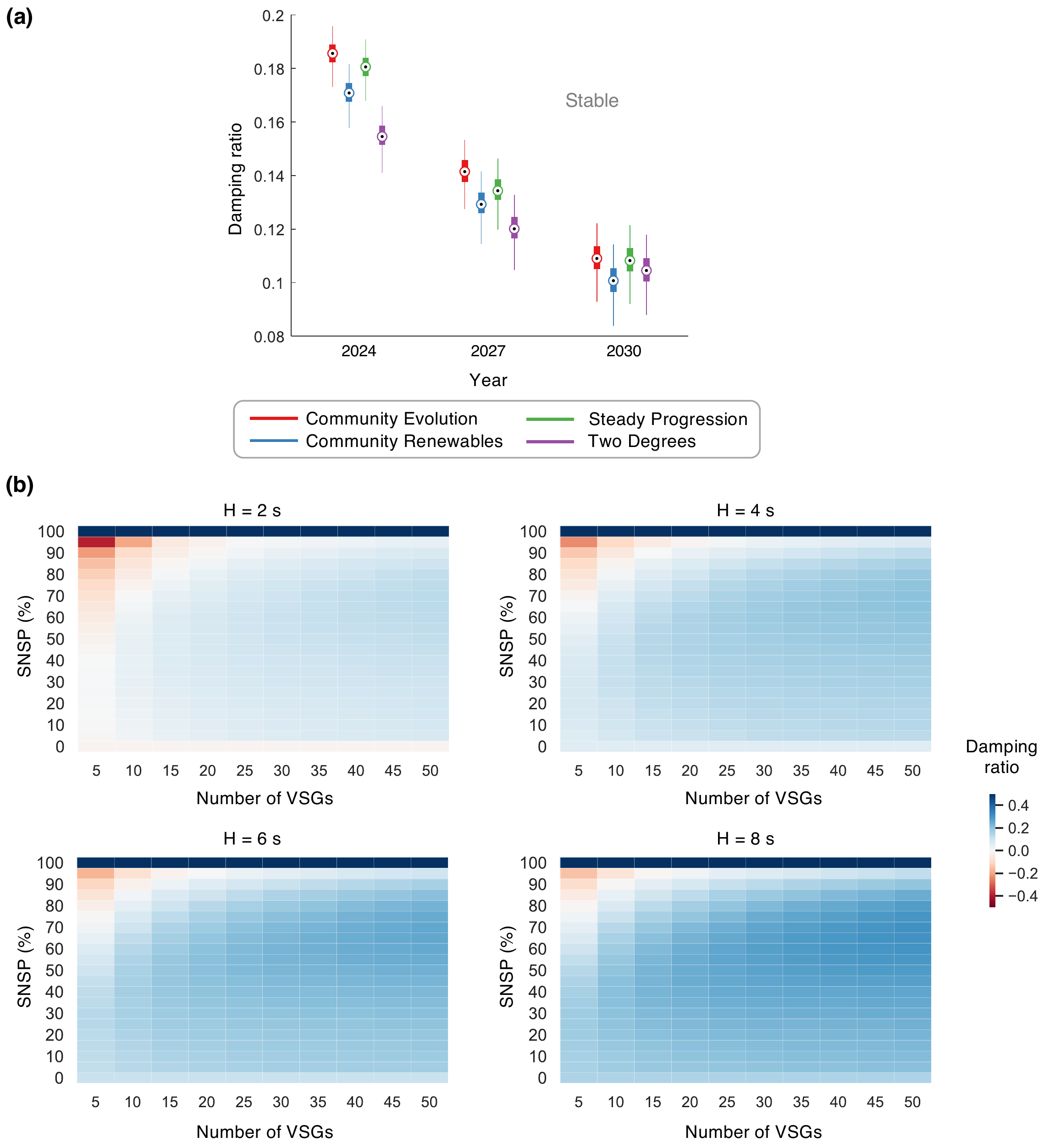}
	\caption{\textbf{Effectiveness of stability storage in the Greater London grid for various energy futures.} 
	How the stability of the projected grids in \textbf{(a)}~Fig.~\ref{fig:Heatmap_variation_fossil_fuel_dependence}(b), and \textbf{(b)}~Fig.~\ref{fig:Heatmap_variation_fossil_fuel_dependence}(c) changes when all VSGs participate as stability stores. The central mark, bottom, and top edges of each box plot in (a) represent the median, 25th, and 75th percentiles, respectively.
	}
	\label{fig:FDC-effectiveness}
\end{figure}

Overall, the stability storage functionality can be leveraged to ensure stability of the Greater London grid for all energy futures shown in Fig.~\ref{fig:Heatmap_variation_fossil_fuel_dependence}(b); see Fig.~\ref{fig:FDC-effectiveness}(a). 
As for the heat maps presented in Fig.~\ref{fig:Heatmap_variation_fossil_fuel_dependence}(c), Fig.~\ref{fig:FDC-effectiveness}(b) shows the same results with stability storage.
While the majority of scenarios now become stable, we comment on the negative damping ratios observed at the top-left.
As was mentioned previously, the instability here arises from the high droop gains of the small-capacity synchronous machines.
Our results show that this cannot be addressed by interventions from the VSGs (along the same lines as external interventions in Fig.~\ref{fig:stable_unstable_scenarios}), and must be necessarily mitigated by phasing out these synchronous generators---this is evident from the top row of the heat maps where the damping ratios are all positive.

\section*{Discussion}
Different grids around the world have different decarbonization pathways and generation mixes. As such, their inertia and SNSP futures are different.
Our study helps energy policymakers and power utilities from diverse power systems to develop a viable roadmap towards achieving higher VSG and therefore renewable penetrations.
To begin, we have shown that there is an optimal number, and an upper limit in the number of grid-forming inverters or VSGs that can be present in the grid while maintaining its small-signal stability.
Additionally, our results demonstrate that recruiting VSGs with smaller capacities jeopardizes the stability of the grid.
Given these constraints, operators can contract prosumers of suitable capacities to participate in forming the grid and providing inertia through the stability market, so that the number VSGs remains below the limit.
Yet, despite such long-term planning by the utilities, the number of VSGs operating can change in the short-term depending on, e.g., generation stochasticity, changing prosumer needs, faults, or other failures. 
Then, the stability storage service proposed in this paper can enable the robust operation, while adequately compensating the prosumers for this additional service.
The prosumers continue to provide stability storage as a stop-gap measure until grid-level coordination obviates its need by calling for an increase in the overall inertia from prosumers, bringing online generators of higher capacity, removing generators with smaller capacities, or effecting updates on primary controller parameters through secondary control.

Finally, our paper has implications on standards development pertaining to inverter-based generation in bulk power systems.
Existing standards such as IEEE 1547~\cite{IEEE1547std} (active standard for distribution grids) and IEEE P2800 (draft standard for transmission grids)~\cite{IEEEP2800std} detail requirements only for grid-following inverters, given that there exist no grid-forming inverters in bulk grids at present.
With the introduction of grid-forming inverters such as VSGs imminent, requirements for these are presently under consideration~\cite{GFMroadmap}, e.g., the GC0137 standard being developed by National Grid in the UK~\cite{GC0137std}.
Our paper directly informs these deliberations by pointing out the need to include stabilizing functions such as variable output impedance in addition to the inertia and droop functions.

\section*{Methods}
\medskip

\noindent \textbf{Power grid specifications.} 
The power grid topology of the Greater London area used in this study was obtained using (i)~electric substation data from National Grid, the UK transmission grid operator~\cite{NationalGridUK}; and (ii)~road network and building data from OpenStreetMap~\cite{OpenStreetMap}.
From the former, we obtain the list of the nine 400~kV/230~kV substations that feed the Greater London county's distribution grid.
From the latter, we obtain the list of buildings in the city, which constitute the loads in the grid.
Subsequently, based on the reasonable assumption that distribution lines are usually laid alongside roads, we derive the power grid topology as a set of spanning trees that connect the nine substations to the loads closest to them; see ref.~\cite{raman2020weaponizing} for more details and SI Appendix, Note~2 for the resultant topology.
Within each subnetwork, the distribution lines are classified into eight different tiers, with the lines closest to the substations or laid alongside larger roadways assigned as Tier-0, and the ones serving the leaf nodes or laid alongside smaller roads assigned as Tier-7.
The values of the line impedances were approximated based on typical R/X ratios and impedance per length values for each tier; see SI Appendix, Note~2.
As for inertia requirements shown in Fig.~\ref{fig:Heatmap_variation_fossil_fuel_dependence}(a), we use projections from National Grid~\cite{Inertia_futures} for the total system inertia (in GVA s) for four future energy scenarios~\cite{FES2019_nationalgrid}, which we divide by the total capacity of synchronous generation (in GVA) obtained from the Government of UK~\cite{DUKES2021_chapter5} to obtain the system inertia constant $H$ (in s).
To obtain the SNSP projections presented in Fig.~\ref{fig:Heatmap_variation_fossil_fuel_dependence}(a), we subtract from 100\% the projected contribution of synchronous generation that we obtain from ref.~\cite{Gomersall_NationalGrid}.
While validation of the distribution grid used for our analyses is clearly not possible due to the unavailability of details of the actual grid, we draw comfort by anchoring the following to reality: (i)~the locations and number of the substations that feed the grid; (ii)~that distribution lines follow the road network; and (iii)~the inertia requirements according to the UK grid operator.
On another note, the substations in our study are represented by generic and equivalent synchronous generators, allowing our analyses to be agnostic to diverse types of generation plants and energy mixes that feed the transmission network.

\medskip
\noindent \textbf{Modeling virtual synchronous generators.}
VSGs are controlled using the conventional real power-frequency, $P-f$, and reactive power-voltage, $Q-V$, droop~\cite{GFMroadmap}.
In more detail, the output frequency $f$ and voltage reference $V_\mathit{ref}$ of each inverter are dependent linearly on the inverter's real power output $P$ and reactive power output $Q$, respectively. 
Formally, the VSG control strategy is defined by the following equations:
\begin{align}
f&=f_0- \frac{k_f}{T_c s+1} P, \\
V_\mathit{ref}&=V_0-\frac{k_v}{T_c s+1} Q \ ,
\label{droop_conv}
\end{align}
where $f_0$ and $V_0$ are the nominal values of the frequency and voltage, $k_f$ and $k_v$ the droop coefficients for frequency and voltage, and $T_c$ the time constant of the power measurement filter.
As more power is drawn from the inverter, its output frequency and voltage deviate from their nominal values to a greater extent, thereby allowing the sharing of power amongst multiple sources in the grid.
Here, the inertia constant $H$ and damping factor $D$ of the inverter in the synchronous machine analogy are respectively:
\begin{align}
H &\equiv \frac{T_c}{2k_f} \\
D &\equiv \frac{1}{k_f} \ .
\end{align}

The overall control scheme for the inverter is cascaded, with the outputs from the droop controller being fed to a voltage controller that tracks the reference $V_\mathit{ref}$, and finally a current controller and sinusoidal pulse-width modulator that generate the switching inputs for the inverter. Detailed control diagrams are presented in SI Appendix, Note~3.
In our simulations, we consider $f_0=50$~Hz. The value of $V_0$ depends on the voltage level of the inverter, i.e., it depends on the tier of the distribution line to which it is connected; see SI Appendix, Note~2. The cut-off frequency $1/(2\pi T_c)$ of the power measurement filter is typically selected between 2-10~Hz~\cite{Mohamed2008adaptive}, taken here as 2~Hz.

\medskip
\noindent \textbf{Stability analysis.} 
We perform stability analysis in the frequency domain using MATLAB.
We do not use time domain simulations as the goal here is to assess the small-signal stability of the grid rather than observing its transient behavior during instabilities.
The transition from conventional generation to fully inverter-based generation is characterized by the SNSP. If $S_d$ is the total power demand in the system and $S_\mathit{sg1}$,$S_\mathit{sg2}$, $\cdots$,$S_\mathit{sgi}$,$\cdots$,$S_\mathit{sgN_{sg}}$ and $S_1$,$S_2$, $\cdots$, $S_j$, $\cdots$, $S_\mathit{N_{inv}}$ are the powers supplied by the synchronous generators $i\in [1,N_{sg}]$ and inverters $j\in [1,N_\mathit{inv}]$, respectively, we define:
\begin{equation}
    SNSP=\frac{\sum_{j=1}^{N_\mathit{inv}}{S_\mathit{j}} }{S_\mathit{total}}
\end{equation}
where $S_\mathit{total}=\sum_{i=1}^{N_\mathit{sg}}{S_\mathit{sgi}}+\sum_{j=1}^{N_\mathit{inv}}{S_\mathit{j}}$.
We now detail how the power system is modeled, and how the stability is assessed for each SNSP value.
For each of the synchronous generators, their relative power capacities $\beta_i$ are modeled to be proportional to the number of buildings connected to that spanning tree:
\begin{equation}
    \beta_i=\frac{\text{Number of buildings connected to spanning tree $i$}}{\text{Total number of buildings}} .
\end{equation}
We assume for simplicity that the inverters are of equal capacity. Then, for a given value of SNSP, considering the total system capacity $S_{total}$ to be constant, the power capacities of the synchronous generators and inverters are calculated as:
\begin{align}
    S_\mathit{sgi}=& \beta_i S_{total} \ (1-SNSP) \ \forall i \in [1,N_\mathit{sg}]    \\
    S_\mathit{j}=& S_{total} \frac{SNSP}{N_{inv}} \ \forall j \in [1,N_\mathit{inv}] 
\end{align}

In this study, we do not simulate the inertia market to determine the inertia contributions by each source. Rather, we consider that the grid operator fixes the overall required inertia, and that each source contributes an inertia proportional to its capacity.
Given the power capacity $S_i$ of a generic generation unit and its inertia constant $H_i$, overall inertia of the system $H_{total}$ is given by:
\begin{equation}
    H_{total}=\frac{\sum_i^{N_{sg}+N_{inv}} H_i S_i}{S_{total}}
\end{equation}
Then, the inertias to be contributed by the $n=N_{sg}+N_{inv}$ sources are:
\begin{equation}
    \begin{bmatrix} H_1 \\ H_2 \\ H_3 \\ \vdots \\ H_n\end{bmatrix}=
    \begin{bmatrix}
    S_1 & S_2 & S_3 & \hdots & S_n\\
    S_2 & -S_1 & 0 & \hdots & 0\\
    S_3 & 0 & -S_1 & \hdots & 0\\
    \vdots & & & \ddots &\\
    S_n & 0 & 0 & \hdots & -S_1
    \end{bmatrix}^{-1}
    \begin{bmatrix} S_{total} H_{total} \\ 0 \\ 0 \\ \vdots \\ 0\end{bmatrix}
\end{equation}
Subsequently, the droop gains of each source are calculated as:
\begin{align}
    k_{fi}&=\frac{T_c}{2 H_i} \\
    k_{vi}&=r_0 k_{fi}
\end{align}
where $r_0$ is the ratio of the voltage droop gain to the frequency droop gain, selected here as 3. Note that these droop gains are expressed in per unit Hz/per unit W or per unit V/per unit VAR, with the per unit conversion done on a common frequency and power base; these are taken as 50~Hz and 10~GVA for our simulations respectively. The voltage base depends on the voltage of the line where the source is situated.

For each iteration of our simulations, the locations of the inverters are randomly selected from nodes that are connected with either Tier-1 or Tier-2 lines. 
More formally, if $\mathcal{E}$ is the set of all edges in the network, the set of potential nodes for locating the inverters is given by $\mathcal{S}= \{ \cup s_i: \mathcal{E}(s_i) \in \{\text{Tier-1}, \text{Tier-2}\}  \}$ where $\mathcal{E}(s_i)$ refers to the set of edges connected to the node $s_i$ in the network. 
Then, the bus admittance matrix of the power network is obtained by eliminating all nodes that do not have an active source (either a VSG or a synchronous generator) through Kron reduction.
The poles of the system are then identified; see SI Appendix, Note~4 for a detailed description. Of the system poles, the complex pole-pair that has the maximum real part is defined as the dominant mode of oscillation; the damping ratio of this pair is used to determine if the grid is stable. We remark here that the system has a null eigenvalue denoting the freedom to shift the reference of all the voltage phasor angles, which is neglected.

Note that for the heat maps in Figs. \ref{fig:Heatmap_variation_fossil_fuel_dependence}(c) and \ref{fig:FDC-effectiveness}(b), when SNSP is 0, the inverters are all removed from the model altogether, and when SNSP is 100\%, the synchronous machines are all removed. Otherwise, the model considering both would produce undefined results as the droop gains of the sources tend to infinity when their power shares tend to zero. This discontinuity in the model results in the observed discontinuity in the damping ratio trends in the first and last rows of the heat maps. 

\medskip

\noindent \textbf{Distributed stability metric.}
We define the following as the distributed stability metric:
\begin{equation}
    \text{DSM}= \frac{\omega_c k_f}{2} \left[ B'-\frac{2 k_v G' G}{1-k_v B} + \frac{k_v G^2 (1-\omega_c k_v B')}{\omega_c (1-k_v B)^2} \right],
\label{eqn:DS_Metric_definition}
\end{equation}
where $\omega_c=\frac{1}{T_c}$ is the cut-off frequency of the power measurement filter in the inverter, representative of the inertia. Further, $G$, and $B$ represent the real and imaginary parts of the static admittance (i.e., the inverse of the impedance) connecting the inverter to a Th\'{e}venin-equivalent of the rest of the grid, respectively.
$G'$ and $B'$ are the real and imaginary parts of the dynamic admittance that captures the transient behavior of the interconnection impedance~\cite{Vorobev2017high}.
From a control theoretic perspective, the DSM reflects the distance of the dominant pole-pair---pertaining to a system where the inverter is connected to an infinite bus that approximates the rest of the grid---from the origin. 
Therefore, the metric for a given VSG is independent from the parameters of the other VSGs in the grid, thereby allowing us to separate the (in)stability contribution of each.
For more details on its derivation, see SI Appendix, Note~5.
Selection of this metric has the following benefits: (i)~it can be calculated locally at each VSG node without any information from the others; and (ii)~it varies intuitively, taking the value of 0 when the VSG is located at an infinite electrical distance from the rest of the grid (i.e., when $G=B=G'=B'=0$, it does not affect the grid stability at all), and monotonically increasing as the electrical distance reduces (i.e., when $G$, $B$, $G'$, and $B'$ increase).
When a VSG contributes stability storage, its effective output impedance to the grid becomes larger and more inductive (which are the conditions for improved stability); the metric decreases and finally saturates at 0, e.g., see Fig.~\ref{fig:FDC_working}(b). 
Finally, we define the incentive/payoff for the prosumer operating the VSG as follows: 0 when the inverter contributes no emulated output impedance, and proportionally increasing with the reduction in the DSM, saturating at a maximum when the metric saturates at 0. 
A VSG participating as a stability store operates as follows.
It continually measures the damping ratio of its power output, e.g., through Prony analysis~\cite{Peng2009comparative}. 
If the damping ratio falls below a preset threshold, here zero, the inverter contributes additional output impedance (see SI Appendix, Note~3 for detailed control diagrams) until its individual instability contribution decreases to a small value, taken in our study as 0.001. 
We remark here that the output impedance should not be increased beyond the value needed to compensate instability, as it would adversely impact the ability of the generator to provide reactive power and respond to dynamic load changes.

\section*{Data availability} 
The power systems data used for the analyses are available at ref.~\cite{website_data}.

\section*{Code availability}
The codes for generating the specific plots are available from the corresponding author upon request.

\small
\bibliographystyle{unsrt}
\bibliography{ref}
\normalsize

\section*{Author contributions}
Gurupraanesh R, Gururaghav R, and J.C.-H.P conceived the study. Gurupraanesh R and Gururaghav R generated the figures and performed the analyses. All authors wrote the manuscript.

\section*{Acknowledgments}
The authors gratefully acknowledge the assistance provided by Dr.~Marcin Waniek, New York University, Abu Dhabi, UAE in generating the map-based plots in the paper. 
The research was conducted at the Singapore-ETH Centre, which was established collaboratively between ETH Zurich and the National Research Foundation Singapore. This research is supported by the National Research Foundation, Prime Minister’s Office, Singapore under its Campus for Research Excellence and Technological Enterprise (CREATE) programme.

\section*{Competing interests}
The authors declare no competing interests.

\section*{Materials and Correspondence}
Correspondence and requests for materials should be addressed to J.C.-H.P.

\end{document}